\documentstyle[prd,aps,preprint,psfig]{revtex}

\begin{document}

\tighten
\draft
\title{Natural Chaotic Inflation in Supergravity}
\author{M. Kawasaki}
\address{Research Center for the Early Universe, University of Tokyo,
  Tokyo, 113-0033, Japan}
\author{Masahide Yamaguchi}
\address{Research Center for the Early Universe, University of Tokyo,
  Tokyo, 113-0033, Japan}
\author{T. Yanagida}
\address{Department of Physics, University of Tokyo, Tokyo 113-0033,
Japan \\ and \\ Research Center for the Early Universe, University of
Tokyo, Tokyo, 113-0033, Japan}

\date{\today}

\maketitle

\begin{abstract}
    We propose a chaotic inflation model in supergravity. In the model
    the K\"ahler potential has a Nambu-Goldstone-like shift symmetry
    of the inflaton chiral multiplet which ensures the flatness of the
    inflaton potential beyond the Planck scale. We show that the
    chaotic inflation naturally takes place by introducing a small
    breaking term of the shift symmetry in the superpotential. This
    may open a new branch of model building for inflationary universe
    in the framework of supergravity.
\end{abstract}



\section{Introduction}

The inflationary expansion of the early universe \cite{inflation} is
the most attractive ingredient in modern cosmology. This is not only
because it naturally solves the longstanding problems in cosmology,
that is the horizon and flatness problems, but also because it
accounts for the origin of density fluctuations \cite{fluc} as
observed by the Comic Background Explorer(COBE) satellite \cite{COBE}.
Among various types of inflation models proposed so far, chaotic
inflation model \cite{chao} is the most attractive since it can
realize an inflationary expansion even in the presence of large
quantum fluctuations at the Planck time. In fact, many authors have
used the chaotic inflation model to discuss a number of interesting
phenomena such as preheating \cite{preheating}, superheavy particle
production \cite{X}, and primordial gravitational waves
\cite{Starobinsky} in the inflationary cosmology \cite{inflation}.

On the other hand, supersymmetry~(SUSY) \cite{SUSY} is widely
discussed as the most interesting candidate for the physics beyond the
standard model since it ensures the stability of the large hierarchy
between the electroweak and the Planck scales against radiative
corrections. This kind of stability is also very important to keep the
flatness of inflaton potential at the quantum level. Therefore, it is
quite natural to consider the inflation model in the framework of
supergravity.

However, the above two ideas, i.e. chaotic inflation and supergravity,
have not been naturally realized simultaneously. The main reason is
that the minimal supergravity potential has an exponential factor,
$\exp(\frac{\varphi_{i}^{\ast}\varphi_{i}}{M_{G}^{2}})$, which
prevents any scalar fields $\varphi_{i}$ from having values larger
than the gravitational scale $M_{G} \simeq 2.4 \times 10^{18}$GeV.
However, the inflaton $\varphi$ is supposed to have a value much
larger than $M_{G}$ at the Planck time to cause the chaotic inflation.
Thus, the above effect makes it very difficult to incorporate the
chaotic inflation in the framework of supergravity. In fact, all of
the existing models \cite{GL,MSYY} for chaotic inflation use rather
specific K\"ahler potential, and one needs a fine tuning in the
K\"ahler potential since there is no symmetry reason for having such
specific forms of K\"ahler potentials. Thus, it is very important to
find a natural chaotic inflation model without any fine tuning.

In this letter, we propose a natural chaotic inflation model where the
form of K\"ahler potential is determined by a symmetry. With this
K\"ahler potential the inflaton $\varphi$ may have a large value
$\varphi \gg M_{G}$ to begin the chaotic inflation. Our models, in
fact, need two small parameters for successful inflation. However, we
emphasize that the smallness of these parameters is justified by
symmetries and hence the model is natural in 't Hooft's sense
\cite{tHooft}.

The existence of a natural chaotic inflation model may open a new
branch of inflation-model building in supergravity, since most of the
model building in supergravity has been concentrated on other types of
inflation models (e.g. hybrid inflation model etc. \cite{LR}).
Furthermore, we consider that future astrophysical observations
\cite{MAP,PLANCK} will be able to select types of inflation models.

Our model is based on the Nambu-Goldstone-like shift symmetry of the
inflaton chiral multiplet $\Phi(x,\theta)$. Namely, we assume that the
K\"ahler potential $K(\Phi,\Phi^{\ast})$ is invariant under the shift
of $\Phi$,
\begin{equation}
  \Phi \rightarrow \Phi + i~C M_{G},
  \label{eq:shift}
\end{equation}
where $C$ is a dimensionless real parameter. Thus, the K\"ahler
potential is a function of $\Phi + \Phi^{\ast}$, $K(\Phi,\Phi^{\ast})
= K(\Phi + \Phi^{\ast})$. It is now clear that the supergravity effect
$e^{K(\Phi + \Phi^{\ast})}$ discussed above does not prevent the
imaginary part of the scalar components of $\Phi$ from having a larger
value than $M_{G}$. We identify it with the inflaton field $\varphi$.
We also stress that the present model overcomes the so-called $\eta$
problem \cite{LR} and it is an alternative to other inflation models
such as D-term inflation models \cite{D-term} and running inflaton
mass models \cite{running}. However, as long as the shift symmetry is
exact, the inflaton $\varphi$ never has a potential and hence it never
causes the inflation. Therefore, we have to introduce a small breaking
term of the shift symmetry in the theory. The simplest choice is to
introduce a small mass term for $\Phi$ in the superpotential,
\begin{equation}
  W = m\Phi^{2}.
  \label{eq:mass0}
\end{equation}
Then, we have the potential,
\begin{equation}
  V = e^{K} \left\{ \left(
      \frac{\partial^2K}{\partial \Phi\partial \Phi^{*}}
    \right)^{-1}D_{\Phi}W D_{\Phi^{*}}W^{*}
    - 3 |W|^{2}\right\},
  \label{eq:potential0}
\end{equation}
with 
\begin{equation}
  D_{\Phi}W = \frac{\partial W}{\partial \Phi} 
    + \frac{\partial K(\Phi+\Phi^{\ast})}{\partial \Phi}W.
  \label{eq:DW}
\end{equation}
Here, $\Phi$ denotes the scalar component of the superfield $\Phi$ and
we have set $M_{G}$ to be unity. We easily see that $V \rightarrow -
\infty$ as $|\varphi| \rightarrow \infty$ with $\Phi+\Phi^{\ast} = 0$
and the chaotic inflation does not take place, where $\varphi = -
i(\Phi-\Phi^{\ast})/\sqrt{2}$.

In this letter, we propose instead the following small mass term in
the superpotential introducing a new chiral multiplet $X(x,\theta)$,
\begin{equation}
  W = mX\Phi.
  \label{eq:mass}
\end{equation}
Notice that the present model possesses $U(1)_{\rm R}$ symmetry under
which
\begin{eqnarray}
    X(\theta) &\rightarrow& e^{-2i\alpha} X(\theta e^{i\alpha}),
    \nonumber \\ 
    \Phi(\theta) &\rightarrow& \Phi(\theta e^{i\alpha}),
  \label{eq:symmetry}
\end{eqnarray}
and $Z_{2}$ symmetry under which
\begin{eqnarray}
    X(\theta) &\rightarrow& - X(\theta e^{i\alpha}), \nonumber \\ 
    \Phi(\theta) &\rightarrow& - \Phi(\theta e^{i\alpha}).
  \label{eq:symmetry2}
\end{eqnarray}
The above superpotential is not invariant under the shift symmetry of
$\Phi$. However, we should stress that the present model is completely
natural in 't Hooft's sense \cite{tHooft}, since we have an enhanced
symmetry (the shift symmetry) in the limit $m \rightarrow 0$. That is,
we consider that the small parameter $m$ is originated from small
breaking of the shift symmetry in a more fundamental theory. We
consider that as long as $m \ll {\cal O}(1)$, the corrections from the
breaking term eq.(\ref{eq:mass}) to the K\"ahler potential are
negligibly small.\footnote{The K\"ahler potential may have also the
induced breaking terms such as $K \simeq |m\Phi|^{2}+\cdots$. However,
these breaking terms are negligible in the present analysis as long as
$|\varphi| \lesssim m^{-1}$.} Then, we assume that the K\"ahler
potential has the shift symmetry eq.(\ref{eq:shift}) and the above
$U(1)_{\rm R} \times Z_{2}$ symmetry neglecting the breaking effects,
\begin{equation}
  K(\Phi, \Phi^{\ast}, X, X^{\ast}) = 
    K[(\Phi + \Phi^{\ast})^{2}, XX^{\ast}]. 
  \label{eq:kahler}
\end{equation}
In the following analysis we take, for simplicity,
\begin{equation}
  K = \frac12 (\Phi + \Phi^{\ast})^{2} + XX^{\ast} + \cdots.
  \label{eq:kahler2}
\end{equation}

\section{Dynamics of inflation}

The Lagrangian density $L(\Phi,X)$ is now given by
\begin{equation}
  L(\Phi,X) = \partial_{\mu}\Phi\partial^{\mu}\Phi^{\ast} 
  + \partial_{\mu}X\partial^{\mu}X^{\ast}
         -V(\Phi,X),
  \label{eq:Lagrangian}
\end{equation}
with the potential $V(\Phi,X)$ given by
\begin{equation}
  V(\Phi,X) = m^{2} e^{K} \left[ ~ 
      |\Phi|^{2}(1+|X|^{4}) 
      + |X|^{2} \left\{ 
      1 - |\Phi|^{2}
      + (\Phi + \Phi^{\ast})^{2}(1+|\Phi|^{2})
      \right\}~\right],
      \label{eq:potential}
\end{equation}     
where we have neglected higher order terms in the K\"ahler potential
eq.(\ref{eq:kahler2}) whose effects will be discussed later.  Here,
$X$ denotes also the scalar component of the superfield $X$.  Now, we
decompose the complex scalar field $\Phi$ into two real scalar fields
as,
\begin{equation}
  \Phi = \frac{1}{\sqrt{2}} (\eta + i \varphi).
  \label{eq:field}
\end{equation}
Then, the Lagrangian density $L(\eta,\varphi,X)$ is given by
\begin{equation}
  L(\eta,\varphi,X) = \frac{1}{2}\partial_{\mu}\eta\partial^{\mu}\eta 
              + \frac{1}{2}\partial_{\mu}\varphi\partial^{\mu}\varphi 
              + \partial_{\mu}X\partial^{\mu}X^{*}
              -V(\eta,\varphi,X),
  \label{eq:Lagrangian2}
\end{equation}
with the potential $V(\eta,\varphi,X)$ given by
\begin{eqnarray}
  \lefteqn{V(\eta,\varphi,X) = m^{2} \exp \left( \eta^{2} + |X|^{2}
                                           \right) } \nonumber \\ 
     &&    \times \left[ ~ 
                 \frac{1}{2} (\eta^{2}+\varphi^{2})(1+|X|^{4}) 
               + |X|^{2} \left\{
                 1 - \frac12(\eta^{2}+\varphi^{2})
               + 2 \eta^{2} \left( 1 + \frac12(\eta^{2}+\varphi^{2}) \right)
                         \right\} ~\right].
  \label{eq:potential2}
\end{eqnarray}     
Note that $\eta$ and $|X|$ should be taken as $|\eta|, |X| \lesssim
{\cal O}(1)$ because of the presence of $e^{K}$ factor. On the other
hand, $\varphi$ can take a value much larger than ${\cal O}(1)$ since
$e^{K}$ does not contain $\varphi$. For $\eta, |X| \ll {\cal O}(1)$,
we can rewrite the potential as
\begin{equation}
  V(\eta,\varphi,X) \simeq \frac{1}{2} m^{2} \varphi^{2} (1+\eta^{2})
  + m^2 |X|^2.
  \label{eq:potential3}
\end{equation}
At around the Planck time, we may have a region \cite{chao} where
\begin{equation}
   \dot{\varphi}^{2} \sim
   (\nabla{\varphi})^{2} \sim V(\varphi) \sim 1.
        \qquad\qquad(\textrm{initial chaotic situation})
\end{equation}  
Here the dot represents the time derivative. In this region the
classical description of the $\varphi$ field dynamics is feasible
because of $|\varphi| \gg {\cal O}(1)$ though quantum fluctuations are
$\delta\varphi \simeq {\cal O}(1)$. Then, as the universe expands, the
potential energy dominates and the universe begins inflation.

Since the initial values of the inflaton $\varphi(0)$ is determined so
that $V(\varphi(0)) \sim \frac{1}{2} m^{2} \varphi(0)^{2} \sim 1$,
$\varphi(0) \sim m^{-1} \gg 1$. (Notice that one has only to demand
$\varphi(0) \gtrsim 15.0$ in order to solve the flatness and horizon
problems. \cite{inflation}) For such large $\varphi$ the effective
mass of $\eta$ becomes much larger than $m$ and hence it quickly
settles down to $\eta = 0$. On the other hand, the $X$ field has a
relatively light mass $m$ and slowly rolls down toward the origin
($X=0$). With $\eta =0$, the potential eq.(\ref{eq:potential3}) is
written as
\begin{equation}
    \label{eq:potential4}
     V(\varphi,X) \simeq  \frac{1}{2} m^{2} \varphi^{2} 
            + m^2 |X|^2.
\end{equation}
Since $\varphi \gg 1$ and $|X| < 1$, the $\varphi$ field dominates the
potential and the chaotic inflation takes place. The Hubble parameter
is given by
\begin{equation}
    H \simeq \frac{m \varphi }{\sqrt{3}}.
\end{equation}

During the inflation both $\varphi$ and $X$ satisfy the slow roll
condition ($|\frac{V''}{V}| \ll 1, \frac12|\frac{V'}{V}|^{2} \ll 1$
where the dash represents the derivative of $\varphi$ or $X$) and
hence the time evolutions are described by
\begin{eqnarray}
    3H \frac{d\varphi}{dt} & \simeq & - m^2 \varphi,\\
    3H \frac{d X}{dt} & \simeq & - m^2 X.
\end{eqnarray}
Here and hereafter, we assume that $X$ is real and positive making use
of the freedom of the phase choice.  From the above equations we
obtain
\begin{equation}
    \left(\frac{X}{X(0)}\right) \simeq  
    \left(\frac{\varphi}{\varphi(0)}\right),
\end{equation}
where $\varphi(0)$ and $X(0)$ are the initial values of $\varphi$ and
$X$ fields. Therefore, $X$ decreases faster than $\varphi$. At the end
of the inflation, i.e. $\varphi \simeq 1$ ($|\frac{V''}{V}| \sim
\frac12|\frac{V'}{V}|^{2} \sim 1$), $X$ is given by
\begin{equation}
    X \lesssim m,
\end{equation}
where we have used $X(0) \lesssim 1$ and $\varphi(0)\sim m^{-1}$. We
see that the $X$ field becomes much smaller than $1$ ($m \sim
10^{-5}$ as shown below). The density fluctuations produced by this
chaotic inflation is estimated as \cite{PS}
\begin{equation}
    \frac{\delta \rho}{\rho}\simeq \frac{1}{5\sqrt{3}\pi}
     \frac{m}{2\sqrt{2}} (\varphi^2 + X^2).
\end{equation}
Since $X \ll \varphi$, the amplitude of the density fluctuations is
determined only by the $\varphi$ field and the normalization at the
COBE scale ($\delta\rho/\rho \simeq 2\times 10^{-5}$ for
$\varphi_{\rm COBE} \simeq 14$ \cite{COBE}) gives \footnote{The
spectral index $n_{s} \simeq 0.96$ for $\varphi_{\rm COBE} \simeq 14$.}
\begin{equation}
    m \simeq 10^{13}~{\rm GeV}.
\end{equation}

After the inflation ends, an inflaton field $\varphi$ begins to
oscillate and its successive decays cause reheating of the universe.
In the present model the reheating takes place efficiently if we
introduce the following superpotential:
\begin{equation}
    W = \lambda X H \bar{H},
    \label{eq:reheat}
\end{equation}
where $H$ and $\bar{H}$ are a pair of Higgs doublets whose $R$-charge
are assumed to be zero and $\lambda$ is a constant.\footnote{$D_{X}W$
is changed to $D_{X}W = (m \Phi + \lambda H \bar{H})(1 + |X|^2)$.
$|H|$ and $|\bar{H}|$ take values $\lesssim {\cal O}(1)$ due to the
factor of $e^{K(H,\bar{H})}$ as the $X$ field. Therefore, the $m \Phi$
term dominates $D_{X}W$ unless $\lambda \gtrsim {\cal O}(1)$, since
$|\Phi(0)| \sim m^{-1}$ at the beginning of the universe, and the
chaotic inflation begins. Once the inflation takes place, $H$ and
$\bar{H}$ acquire masses of the order the Hubble scale and rapidly go
to zero. Thus, the above superpotential eq.(\ref{eq:reheat}) does not
affect the dynamics of the inflation.} Then, we have the coupling of
the inflaton $\varphi$ to the Higgs doublets as
\begin{equation}
    L \sim  \lambda m \varphi H \bar{H},
\end{equation}
which gives the reheating temperature 
\begin{equation}
    T_R \sim 10^{9}~{\rm GeV} 
    \left(\frac{\lambda}{10^{-5}}\right)
    \left(\frac{m}{10^{13}{\rm GeV}}\right)^{1/2}.
\end{equation}
In order to avoid the overproduction of gravitinos, the reheating
temperature $T_{R}$ must be lower than $10^{9}$GeV \cite{Ellis}, which
requires the small coupling $\lambda \lesssim 10^{-5}$. The small
coupling $\lambda$ is naturally understood in 't Hooft's sense
\cite{tHooft} provided that $H\bar{H}$ is even under the $Z_{2}$
symmetry in eq.(\ref{eq:symmetry2}).

So far we have taken the minimal K\"ahler potential and neglected
higher order terms like $(\Phi + \Phi^{*})^4, |X|^4$, and $\cdots$.
Here, we make a comment on the higher terms in the K\"ahler potential.
Since the leading quadratic terms make the expectation values of
$\eta$ and $X$ fields less than 1, the inflation dynamics is almost
unchanged in the presence of the higher terms. The only relevant
difference comes from the $ \zeta |X|^4$ ( $\zeta$: constant) term
which induces the effective mass of $X$ given by
\begin{equation}
    m_{X}^2 = 2m^2 - 2\zeta m^2 \varphi^2 \simeq - 2\zeta  m^2 \varphi^2. 
\end{equation}
Thus, $\zeta$ should be negative to ensure the positiveness of
$m_{X}^2$. If $|\zeta| \gtrsim 1$, the effective mass becomes larger
than the Hubble parameter and the $X$ quickly settles down to $X=0$
without slow-roll.

\section{Conclusion}

We have shown that a chaotic inflation naturally takes place if we
assume that the K\"ahler potential has the Nambu-Goldstone-like shift
symmetry of the inflaton chiral multiplet $\Phi$ and introduce a small
breaking term of the shift symmetry in the superpotential
eq.(\ref{eq:mass}). Unlike other inflation models the chaotic
inflation model has no initial value problem and hence it is the most
attractive. However, it had been difficult to construct a natural
chaotic inflation model in the framework of supergravity because the
supergravity potential generally becomes very steep beyond the Planck
scale. Therefore, the existence of a natural chaotic inflation model
may open a new branch of inflation-model building in supergravity.
Furthermore, the chaotic inflation is known to produce gravitational
waves ( tensor metric perturbations )~\cite{Starobinsky} which might
be detectable in future astrophysical observations~\cite{MAP,PLANCK}.

\subsection*{Acknowledgments}

M.Y. is grateful to J. Yokoyama for useful discussions. M.K. and T.Y.
are supported in part by the Grant-in-Aid, Priority Area
``Supersymmetry and Unified Theory of Elementary Particles''(\#707).
M.Y. is partially supported by the Japanese Society for the Promotion
of Science.

\end{document}